\documentclass{desyproc}

\begin{document}

% %%%%%%%%%%%%%%%%%%%%%%%%%%%%%%%%%%%%%%%%%%%%%%%%%%%%%%%%%%%%%%%%%%%%%%%%%%%%%%%%
%   Title of your contribution
% %%%%%%%%%%%%%%%%%%%%%%%%%%%%%%%%%%%%%%%%%%%%%%%%%%%%%%%%%%%%%%%%%%%%%%%%%%%%%%%%

\title{ 
%\hspace*{2.2 cm }  
Precision DIS data and TMD parton distributions\\ 
}

% %%%%%%%%%%%%%%%%%%%%%%%%%%%%%%%%%%%%%%%%%%%%%%%%%%%%%%%%%%%%%%%%%%%%%%%%%%%%%%%%

% %%%%%%%%%%%%%%%%%%%%%%%%%%%%%%%%%%%%%%%%%%%%%%%%%%%%%%%%%%%%%%%%%%%%%%%%%%%%%%%%
%   Authors
% %%%%%%%%%%%%%%%%%%%%%%%%%%%%%%%%%%%%%%%%%%%%%%%%%%%%%%%%%%%%%%%%%%%%%%%%%%%%%%%%

%
%   For the author list please adhere to the format of one of the following
%   three examples:
%
%

     \author{{\slshape F.~Hautmann$^{1,2,3}$ and  H.~Jung$^{4,5}$}\\[1ex]
        $^1$Rutherford Appleton Laboratory\\
        $^2$Physics and Astronomy, University of Southampton\\
        $^3$Theoretical Physics, University of Oxford\\   
        $^4$Deutsches Elektronen Synchrotron\\
        $^5$Particle Physics, University of Antwerp}
%
%   * use the following for an author speaking on behalf of a collaboration
%
%     \author{{\slshape Joe Smith}  for the FOO Collaboration\\[1ex]
%     CERN, 1211 Gen\`eve 23, Switzerland}
%

%\author{{\slshape Joe Smith}  for the FOO Collaboration\\[1ex]
%CERN, 1211 Gen\`eve 23, Switzerland}

% %%%%%%%%%%%%%%%%%%%%%%%%%%%%%%%%%%%%%%%%%%%%%%%%%%%%%%%%%%%%%%%%%%%%%%%%%%%%%%%%

\contribID{ZZ}
\confID{UU}
\desyproc{DESY-PROC-2012-YY}
%\doi
\maketitle

% %%%%%%%%%%%%%%%%%%%%%%%%%%%%%%%%%%%%%%%%%%%%%%%%%%%%%%%%%%%%%%%%%%%%%%%%%%%%%%%%
%   Abstract
% %%%%%%%%%%%%%%%%%%%%%%%%%%%%%%%%%%%%%%%%%%%%%%%%%%%%%%%%%%%%%%%%%%%%%%%%%%%%%%%%

\begin{abstract}
General formulations of QCD factorization for 
hadronic collisions extend the notion of ordinary 
 parton distributions 
to  transverse-momentum dependent  (TMD)      
 parton  density and parton decay  
functions.   We discuss the use   of 
the recent 
high-precision  deep-inelastic scattering (DIS)  
measurements  for determination of  
 TMD distributions. 
These are relevant for both  
low-$p_T$ and 
high-$p_T$ physics in hadron collisions.  We 
comment on    
  applications to multi-jet final states 
associated with  electroweak 
gauge boson production at the LHC.

\vskip 0.3cm 
\hspace*{0.1cm} Based on talks given  at  the conferences   
{\it  QCD at Cosmic Energies - VI} (Paris, May 2013),   \\    
\hspace*{0.3cm}  
{\it  MPI-TAU} (Tel Aviv, October 2012),  
 {\it  Nonperturbative QCD 2011} (Paris, June 2011)

\end{abstract}

% %%%%%%%%%%%%%%%%%%%%%%%%%%%%%%%%%%%%%%%%%%%%%%%%%%%%%%%%%%%%%%%%%%%%%%%%%%%%%%%%

% %%%%%%%%%%%%%%%%%%%%%%%%%%%%%%%%%%%%%%%%%%%%%%%%%%%%%%%%%%%%%%%%%%%%%%%%%%%%%%%%
%   Contents
% %%%%%%%%%%%%%%%%%%%%%%%%%%%%%%%%%%%%%%%%%%%%%%%%%%%%%%%%%%%%%%%%%%%%%%%%%%%%%%%%

\vskip 0.8 cm

\section{Introduction}

Transverse momentum dependent (TMD) parton distributions 
extend the concept of ordinary (integrated) parton distribution 
functions (pdfs) to include transverse momentum and polarization 
degrees of freedom. 
They  encode nonperturbative information on hadron structure  
which is  essential  in the context of QCD factorization theorems 
for   multi-scale, non-inclusive   collider  observables. 
Classic examples are found in   processes  such as  Drell-Yan (DY) 
production when 
the vector boson invariant mass is large compared to its transverse 
momentum,  and deeply inelastic scattering (DIS)  when  the photon-hadron 
center-of-mass energy is large compared to the photon virtuality. 
In each of these cases, TMD pdfs obey  
evolution equations which generalize 
renormalization group evolution to the appropriate multi-scale 
regime.  In each case   such generalized evolution equations, 
once combined with  factorization of the physical cross section in terms of 
TMD pdfs, allow one to resum logarithmically enhanced contributions 
to the perturbation series expansions   for the  physical observables 
to  all orders in the   QCD coupling.  

A general  program for   TMD pdfs  phenomenology  
was proposed in~\cite{mert-rog}.  
Recently, the  TMDlib library has  been 
started~\cite{tmdlibpaper}  to  provide a platform 
for  the collection and comparison   of 
phenomenological  studies of TMD pdfs.  This aims to incorporate  
as  broad as possible a set of processes and collect  results 
from different  approaches to TMD evolution and fitting.  
This includes e.g. parton-model 
fits~\cite{anselmino,signori} of low-energy 
 data~\cite{hermes13,compass13}; 
CSS~\cite{css85,jcc-book}  fits~\cite{jccetal12,landry,guzzi} 
of  DY data~\cite{d0-11,atlas-11};  
TMD fits~\cite{herafitter1410,hj-updfs-1312,hj1206} 
of high-energy  DIS 
data~\cite{aaron2009,Aaron:2009aa,comb-charm}. 

This  article     focuses    on the use of the precision DIS data 
from the  combined HERA 
measurements~\cite{Aaron:2009aa,comb-charm}   
to make a determination of 
the TMD gluon 
density~\cite{hj-updfs-1312} 
including, for 
the first time, experimental and theoretical 
 uncertainties. 
The basis of this determination is the factorization of TMD 
pdfs in DIS at high energy. We briefly recall the   main elements  of this 
 in Sec.~2.  The results of the fits to precision 
measurements are summarized  in Sec.~3.  We  discuss 
implications and give final comments  in Sec.~4.

\section{DIS at high energy  and factorization of TMD pdfs}
\label{sec:disx}

\begin{figure}[htb]
\begin{center} 
 \includegraphics[scale=0.42]{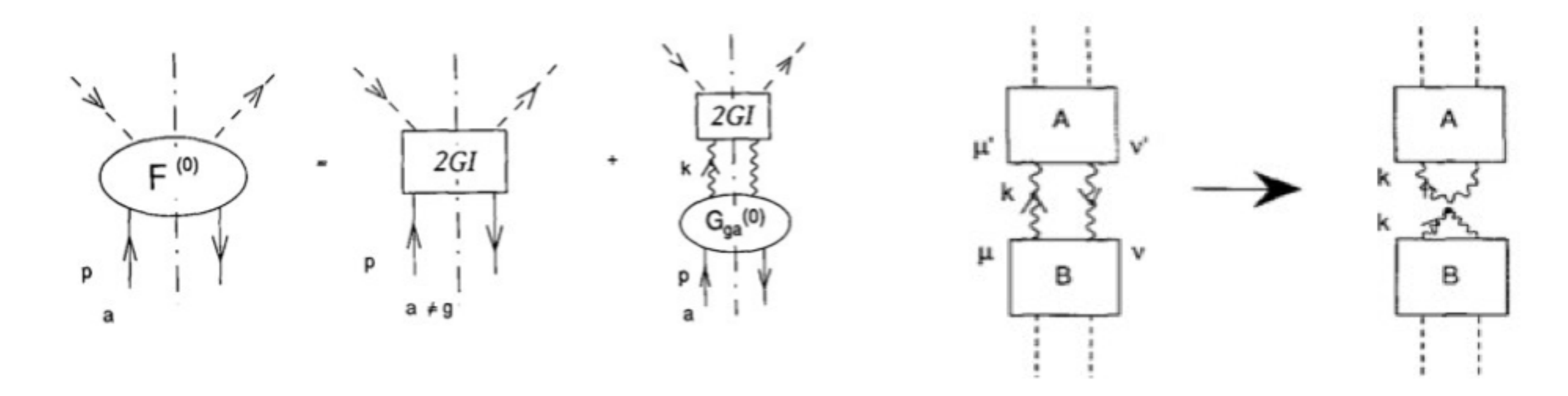}
  \caption{\it (left) Transverse momentum dependent 
high-energy factorization for DIS structure functions; 
(right) decoupling of 2GI kernels via the high-energy projector.  
}
\label{Fig:hefpictdis}
\end{center}
\end{figure}

Consider deeply inelastic lepton-hadron scattering 
in the high-energy region 
 $ s \equiv 2 q \cdot p  \gg Q^2$, where $p$ is the 
hadron four-momentum, $q$ is the 
four-momentum transferred by the lepton, and 
$Q^2 = - q^2$. In this region the perturbation  
series expansions for the DIS structure 
functions $F_j$ ($ j = 2 , L$)  are affected by 
potentially large  logarithms $ 
( \alpha_s \ln s / Q^2 )^n $ to all orders in the 
QCD coupling $\alpha_s$. 
TMD high-energy factorization for 
the structure functions~\cite{ch94}   
is pictured in Fig.~\ref{Fig:hefpictdis}. It is given in 
terms of two-gluon irreducible (2GI),  
perturbatively-calculable kernels  and gluon 
 (off-shell)  Green's functions. The TMD gluon 
density is defined from the latter by 
 using the high-energy projector $P_H$~\cite{ch94,hef} 
over spin and momentum. 

The TMD factorization 
(Fig.~\ref{Fig:hefpictdis})  allows  one to identify and 
sum the logarithmically enhanced corrections to 
coefficient functions and anomalous dimensions 
for $ x =  Q^2 / (2 p\cdot q)  \ll 1 $,  
fully taking into account the factorization scheme and 
factorization scale dependence. 
This has been applied to leading and next-to-leading 
logarithmic accuracy~\cite{ch94,fl98,dis-x-phen}.

The 
perturbation expansions in the flavor-singlet 
sector  are single-logarithmic at high energy. 
For example, the gluonic 
hard-scattering coefficient function  $
C_{ 2 }^g (N, \alpha_s , Q^2/ \mu^2 )  $  
for the DIS structure  function $F_2$, as a function of the 
moment $N$ Mellin-conjugated to $x$, has the 
$\alpha_s$ expansion 
at scale $\mu^2 = Q^2$~\cite{ch94} 
\begin{eqnarray} 
 &&   C_{ 2 ,  N}^g (\alpha_s , Q^2/ \mu^2 = 1  ) 
\nonumber\\ 
&& 
 =  { \alpha_s \over { 2 \pi}  }  T_R N_f {2 \over 3} 
\left\{  1 + 1.49 \  {  {\overline \alpha}_s \over N }  
+ 9.71 \left(  {  {\overline \alpha}_s \over N } \right)^2  
+ 16.43  \left(  {  {\overline \alpha}_s \over N } \right)^3 + 
{ \cal O }      \left(  {   \alpha_s \over N } \right)^4  
\right\}  \;  , 
\nonumber
\end{eqnarray} 
 where $  {\overline \alpha}_s  =  \alpha_s  C_A / \pi$, 
$C_A = 3$, $T_R = 1 / 2$, and  the coefficients are given 
   in the   
$ {\overline{\rm MS}} $ minimal subtraction  scheme.  
The $N \to 0$ poles $ \alpha_s 
  (     \alpha_s / N  )^k $, $ k \geq 1$,  
correspond in $x$-space 
 to next-to-leading-logarithmic  (NLL) 
higher-loop corrections $ \alpha_s^2   
 (\alpha_s  \ln  x )^{k-1}  $. With respect 
to the first two terms 
 (one-loop~\cite{cfp}    and 
two-loop~\cite{vannee}, corresponding to leading order 
(LO) and next-to-leading order (NLO)),   
the next two terms (i.e.,  
three-loop~\footnote{The three-loop coefficient 
  agrees with  the  complete  
next-to-next-to-leading-order (NNLO) 
calculation~\cite{mochetal}.} and four-loop)   are 
logarithmically enhanced. Moreover, their   
 numerical  coefficients are significantly larger than 
the LO and NLO ones.   The physical origin of the logarithmically enhanced higher-loop terms 
  lies  with  contributions   
to   QCD  multi-parton matrix elements
 from  regions  which 
are not ordered in the 
initial-state transverse momenta.

 Given these results, there is little 
theoretical justification for   treating 
the region  $x \ll 1$ by 
truncating    the perturbative 
 expansion to  fixed NLO (or NNLO) 
 level. In this region 
fixed-order perturbative approaches,  
however successful  
phenomenologically, are    
theoretically disfavored. 
   (See~\cite{avogt12} for 
discussion of double-logarithmic 
corrections to DIS in the region   $x \to 1$ 
 and to timelike 
processes,  and  the  need for resummations in 
 these cases.)        
TMD factorization is  required to treat 
the physics of the scaling violation in the 
high-energy 
regime,  where transverse momentum ordering in the 
partonic  initial state  does not apply. 

This  is the motivation for the work we are going to  describe  
in the next section, in which a quantitative step is 
taken toward going beyond fixed-order phenomenology, 
and using the high-precision combined 
measurements~\cite{Aaron:2009aa,comb-charm}
for determination of the nonperturbative TMD gluon density function. 

It is worth noting that similar motivation is at the     
basis of  approaches such as those   
in~\cite{dis-x-phen}. These works go beyond 
fixed-order  analyses  by including 
perturbative resummations. However, 
they do not  attempt to extract information on 
the  TMD gluon density from  experimental data.

\section{TMDs  from the  combined HERA data}
\label{sec:fits}

The    combined measurements  of  proton's DIS  structure functions 
  at the  HERA  collider~\cite{Aaron:2009aa}   
provide high-precision data 
 capable of constraining  parton density functions 
 over a  broad   range of the kinematic variables. These  data 
have been widely used for determinations of the 
integrated (collinear)  pdfs  and for  applications  of 
these pdfs to 
   LHC processes~\cite{herafitter1410,pdf4lhc-alekhin}.  
Ref.~\cite{hj-updfs-1312}  takes a first step to 
extend this approach by studying 
  what can be learnt
 from the precision  DIS data  about  hadron  
structure beyond collinear level, 
using the TMD factorization of Sec.~2.

Phenomenological applications of this approach 
at HERA collider energies 
 require   matching  of  $x \ll 1 $ contributions with 
finite-$x$ contributions. 
To this end, in~\cite{hj-updfs-1312}    the evolution of the gluon density 
 is obtained by combining the resummation of  
small-$x$ logarithmic contributions~\cite{lipatov}   
with medium-$x$ and large-$x$ contributions to parton splitting~\cite{dglaprefs}
according to the CCFM evolution equations~\cite{skewang}. 
This     is done via   the  exclusive 
parton-branching  Monte Carlo 
implementation~\cite{ref_updfevolv}   
of   CCFM  evolution.

\begin{figure}[htb]
\begin{center} 
 \includegraphics[scale=0.26]{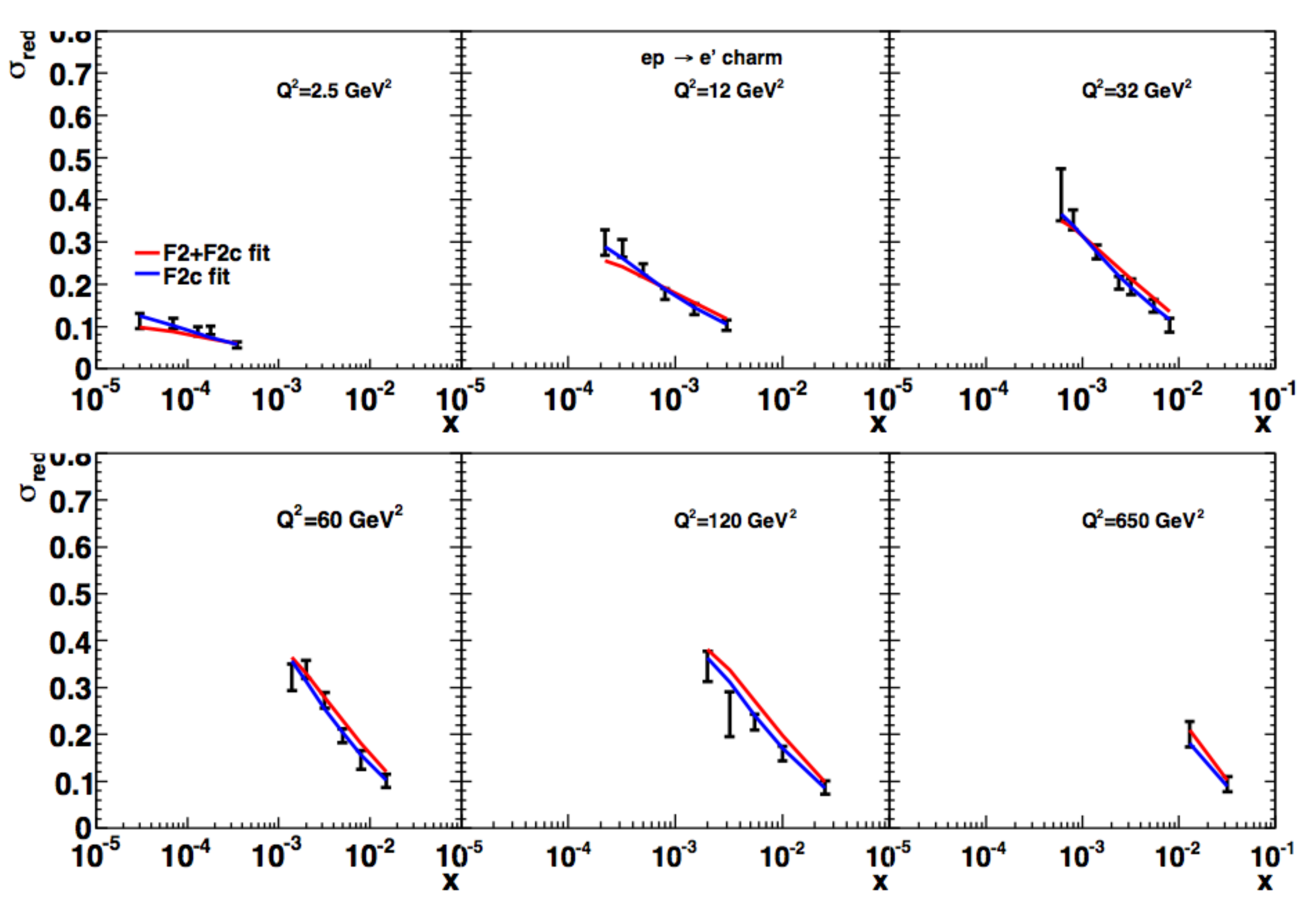}
 \includegraphics[scale=0.26]{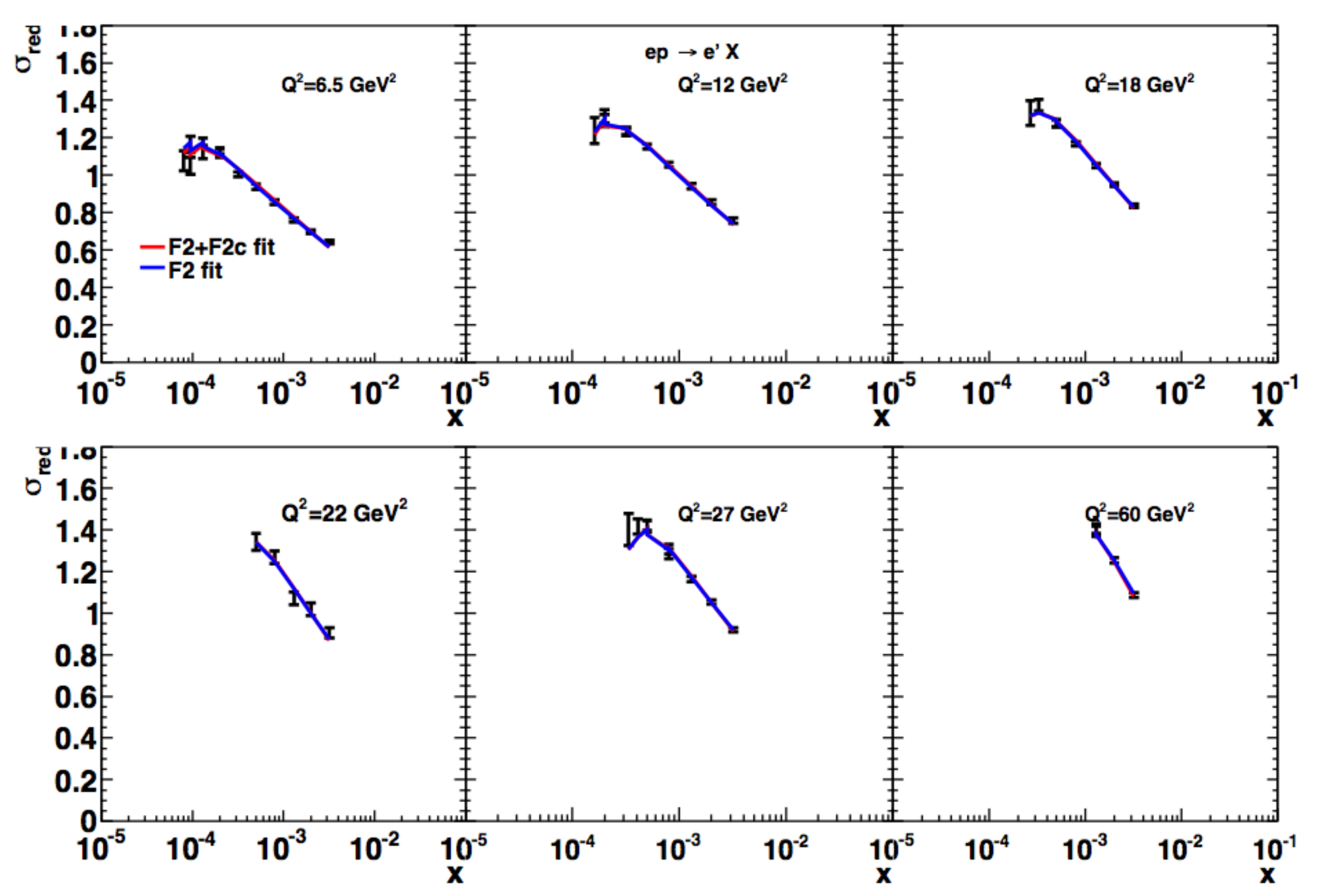}
  \caption{\it The fit~\protect\cite{hj-updfs-1312}  to 
DIS  precision     measurements: 
(top) charm leptoproduction 
data \protect\cite{comb-charm};  
(bottom) inclusive structure function 
data  \protect\cite{Aaron:2009aa}.
}
\label{fig:f2datafit}
\end{center}
\end{figure}

The  TMD gluon distribution at  the initial scale 
  of  the evolution 
 is  determined from fits to precision DIS data.   
Fits are performed to the combined HERA 
charm-quark leptoproduction 
data~\cite{comb-charm} over the whole kinematic range of the 
measurement, corresponding to $Q^2 > 2.5$~GeV$^2$,  
and  to  the  $F_2$ structure function 
data~\cite{Aaron:2009aa} in the  range 
$Q^2 > {\overline Q}^2 $, $ x < {\overline x}$,  
where we take $ {\overline Q}^2  = 5 $ GeV$^2$, 
$ {\overline x} = 5 \cdot 10^{-3}$.\footnote{The  
 restriction  on the kinematic range for  the $F_2$  data 
is    motivated 
by    the fact that the approach relies    on  
 perturbative factorization theorems,  which 
classify higher-order  corrections 
 according to the logarithmic hierarchy  
based on  high  $Q^2$ and low  $x$.   The 
choice of the cuts $ {\overline Q}$, $ {\overline x}$  presented here 
may however be regarded as   conservative.  
As discussed in~\cite{hj-updfs-1312},   extensions to 
lower $Q^2$ and higher $x$ are 
possible ---  this  may be part of  
 programs  for investigating 
 future  uses of HERA precision  
measurements~\cite{futurehera2014}. See e.g.~\cite{hs07} for 
some of the issues involved  in approaches to low $Q^2$ using 
unintegrated parton correlation functions.}   
The fits to the HERA  measurements 
are obtained with  
the \verb+herafitter+  
package~\cite{herafitter1410,jamesroos,aaron2009,Aaron:2009aa},    
 treating the correlated systematic uncertainties 
separately from the uncorrelated statistical and systematic uncertainties.  

\begin{table}[ht]
\centering  
\begin{tabular}{|c|c|c|c|}
\hline
       	&$\chi^2 / ndf  ( F_2^{({\rm{charm}})}  ) $	&$ 
\chi^2 / ndf  ( F_2 ) $	& $\chi^2 / ndf  \left( F_2  \mbox{ and } F_2^{({\rm{charm}})}  \right) $	  \\ \hline
3-parameter 		&0.63 	&1.18  &	1.43  \\ \hline
5-parameter		&0.65	&1.16  &	1.41	   \\ \hline
\end{tabular}	
\caption{\it The values of 
$\chi^2 / ndf $~\protect\cite{hj-updfs-1312}  
corresponding to the 
 best fit  
for charm structure function  $F_2^{({\rm{charm}})} $,   
 for inclusive structure function $F_2$,  and for the 
 combination of $F_2^{({\rm{charm}})}$ and  $F_2$.}
\label{tablechi}
\end{table}

The Monte Carlo implementation~\cite{ref_updfevolv} of 
 CCFM evolution 
includes 
two-loop running coupling, finite-$x$ gluon splitting  
and  energy-momentum consistency 
 constraint~\cite{hj-updfs-1312,hjdis1314}. 
In addition to the  gluon-induced process 
$\gamma^* g^* \to q\bar{q}$ the contribution 
from valence quarks is included 
via $\gamma^* q \to q$    
by using a  CCFM evolution of valence 
quarks~\cite{Deak:2010gk}.\footnote{TMD sea quarks~\cite{hent12}   
 are not yet included.} 
Two different functional forms (with three and five parameters) 
for the starting gluon distribution 
are used.  
The corresponding results  
for the values of $\chi^2$ per 
degree of freedom   are shown in 
Table~\ref{tablechi}. This    reports the  $\chi^2$ per 
degree of freedom for the best fit to the charm 
structure function $F_2^{({\rm{charm}})} $~\cite{comb-charm},   
 the inclusive structure function 
$F_2$~\cite{Aaron:2009aa},   and  a combination of both. 
Fig.~\ref{fig:f2datafit} shows the description of the charm 
leptoproduction and inclusive structure function measurements, 
by the individual fits and a combined fit. Plotted are the reduced cross 
sections defined in~\cite{Aaron:2009aa,comb-charm}.

\begin{figure}[htb]
\begin{center} 
 \includegraphics[scale=0.32]{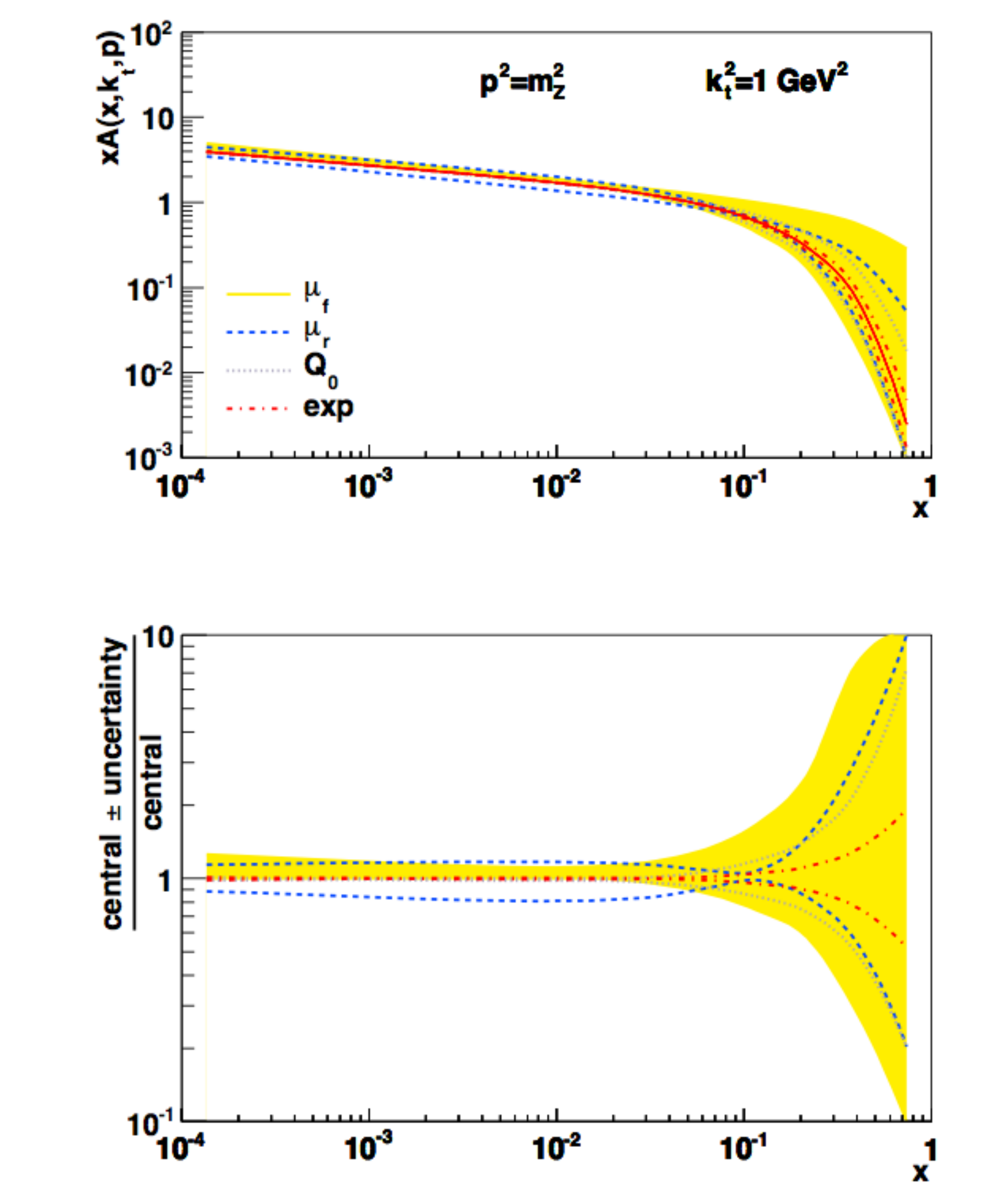}
 \includegraphics[scale=0.32]{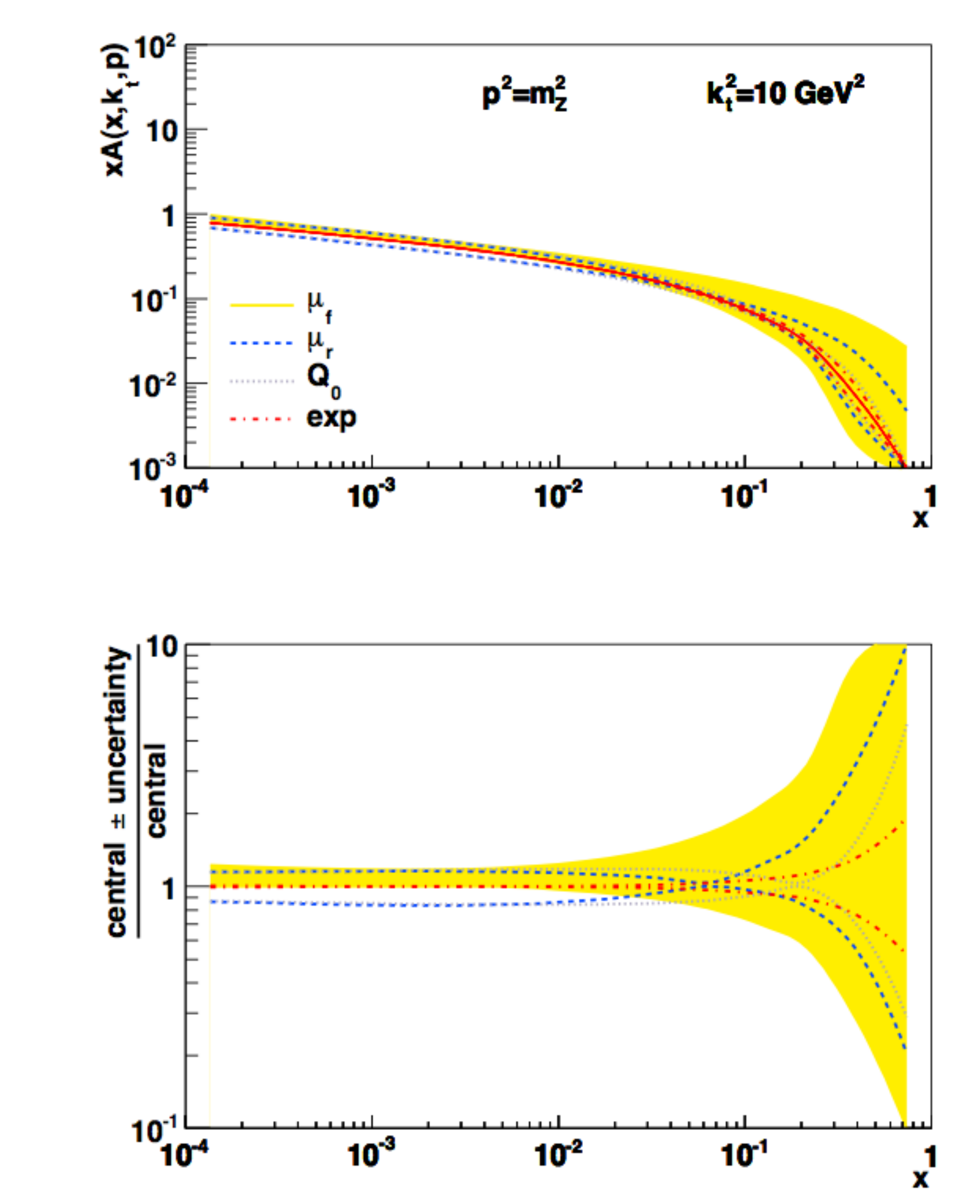}
  \caption{\it Experimental and theoretical 
uncertainties of the unintegrated TMD gluon density versus $x$ 
for different values of transverse momentum 
at  $p^2 = m_Z^2$~\protect\cite{hj-updfs-1312}. The 
yellow band gives the uncertainty 
from the factorization scale variation, 
while  the  curves indicate  
the uncertainties from renormalization scale, starting scale, and 
experimental errors, as   described 
in~\protect\cite{hj-updfs-1312}. 
}
\label{Fig:updf-uncertainty}
\end{center}
\end{figure}

We see from Table~\ref{tablechi} that 
the best-fit  $\chi^2 / ndf $  is below 1 for  the charm 
structure function, while it is around 1.18 for the inclusive 
structure function.   This 
is in accord with  expectations based on   
   charm production coupling 
 predominantly to gluons  
while the inclusive structure function couples 
to quark channels,  for which sea quark distributions are not yet 
included  at TMD level. 
Despite the restricted kinematic range of the 
experimental data analyzed, 
the great  precision  of the measurements 
provides a highly  nontrivial test  of the approach.   
Based on the  fit  to  charm-quark and inclusive 
data, Ref.~\cite{hj-updfs-1312} presents two sets 
of TMD pdfs,   JH-2013. It also 
presents    experimental and theoretical 
uncertainties associated with these.  An example is  
shown  in Fig.~\ref{Fig:updf-uncertainty}, where 
different contributions to the uncertainty (both from 
experiment and from theory)  are plotted 
for different values of transverse momentum 
at evolution scale equal to the $Z$ boson mass.  
The uncertainties are small for low $x$ values, while 
they become sizeable 
for high $x$ values, where there is little constraint 
 from  experimental  data, and the 
theoretical accuracy  of the calculation   decreases.

\section{Discussion}   
\label{sec:disc}

The work  described in Sec.~3  is 
the  first determination of  the
TMD gluon distribution  
which includes 
the precision DIS  measurements and  
which provides  
   experimental and theoretical    uncertainties. 
As discussed earlier,  despite  
the limited  kinematic range of the $F_2$ data,    
$Q^2 > 5  $ GeV$^2$, $ x < 5 \cdot  10^{-3}$,  
their high precision implies a 
stringent test of the approach based on 
TMD  factorization at high energy. 

The method~\cite{hj-updfs-1312} for determining 
uncertainties associated with the TMD pdfs,  
implemented within 
 \verb+herafitter+~\cite{herafitter1410}, 
includes  theoretical  uncertainties on the 
TMD gluon density  from  variation of 
the factorization scale  and renormalization scale.  
This  differs from that usually 
 followed in 
determinations of ordinary, collinear 
 pdfs from   fixed-order perturbative 
treatments.  In this case, no uncertainty on the pdfs is considered from 
scale variation. It is only 
 when computing predictions for 
any specific observable 
that  the theoretical uncertainty on  the 
predictions  is estimated by   scale variation.  In 
the  approach~\cite{hj-updfs-1312}   one is 
 interested to  study  the uncertainty 
from varying scales in the theoretical calculation  used  
to determine the pdf.  
For example  in Fig.~\ref{Fig:updf-uncertainty}  
  the 
renormalization scale (blue dashed curves) and 
  the factorization 
scale  (yellow band)  are varied by  a factor of 2 above 
and below the central value.  
It can  be   interesting  to extend this approach to 
collinear fits.  

Having examined physics in the HERA region, 
one can ask whether this theoretical framework 
can be used  to treat  higher $p_T$ processes. 
 To this end observe that, unlike 
 forms of TMD factorization specifically designed  
for the  low-$p_T$ region~\cite{css85,scet12,mulders},  
\begin{figure}[htb]
\includegraphics[scale=0.4]{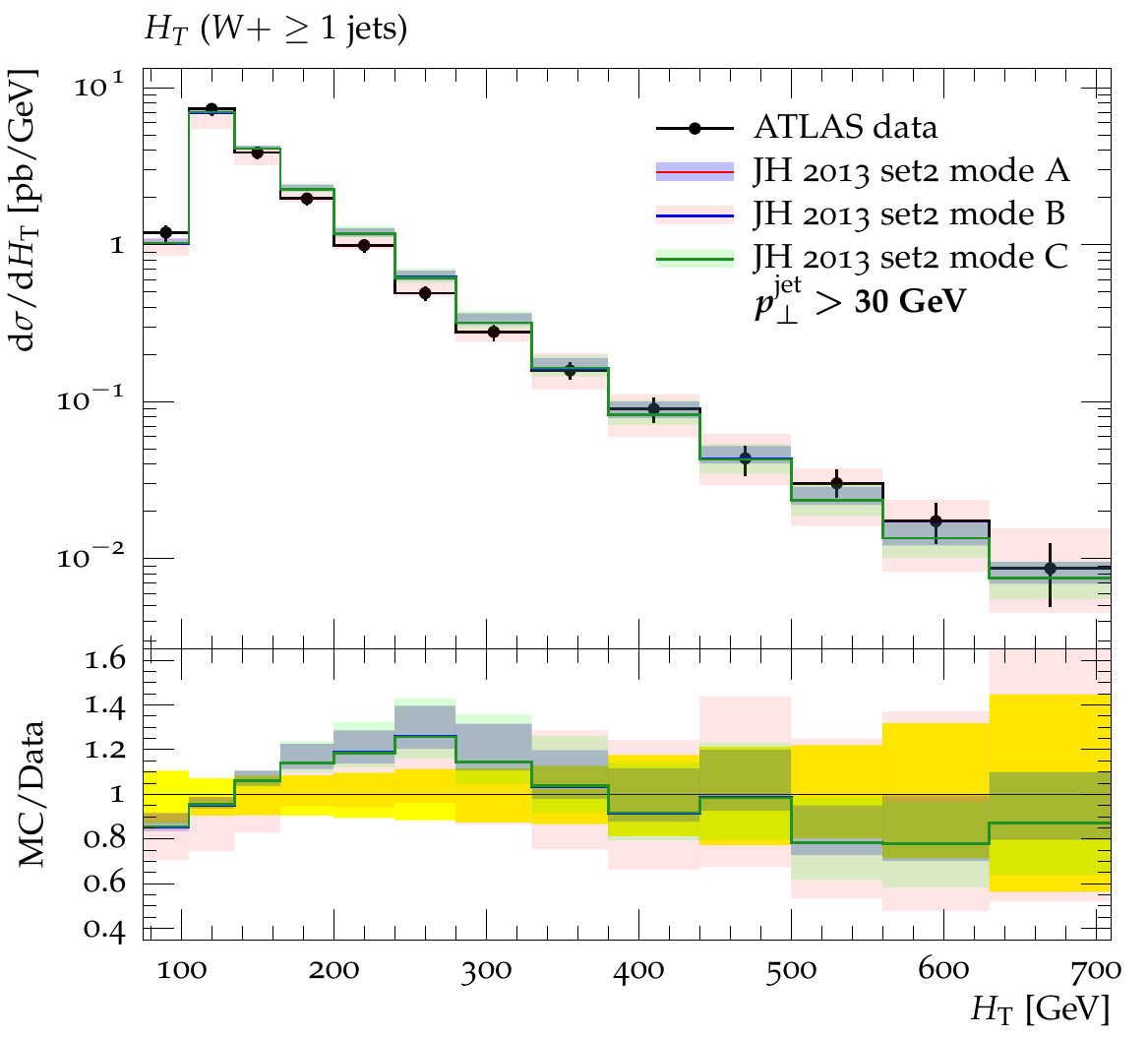}
\includegraphics[scale=0.4]{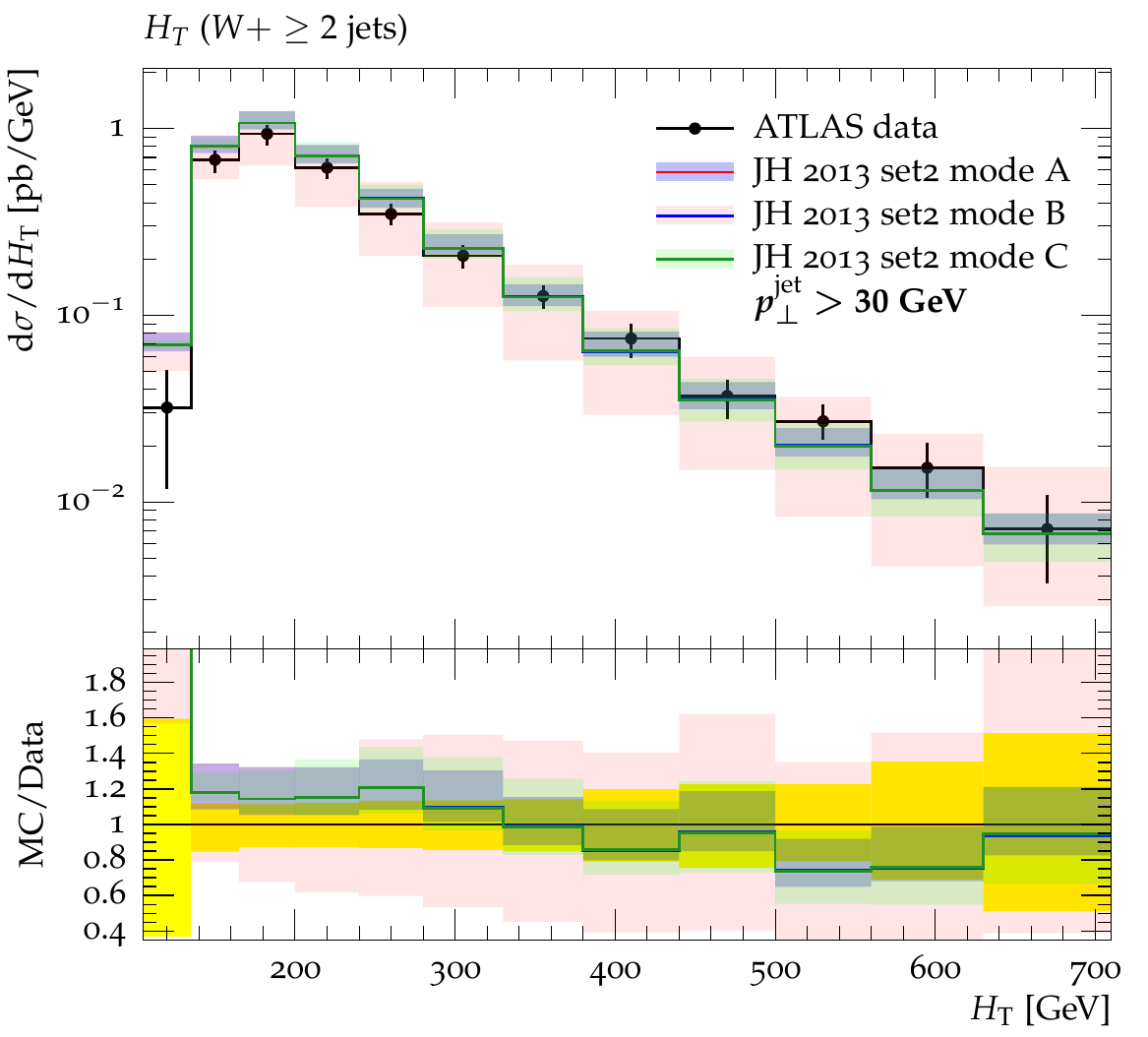}
\includegraphics[scale=0.4]{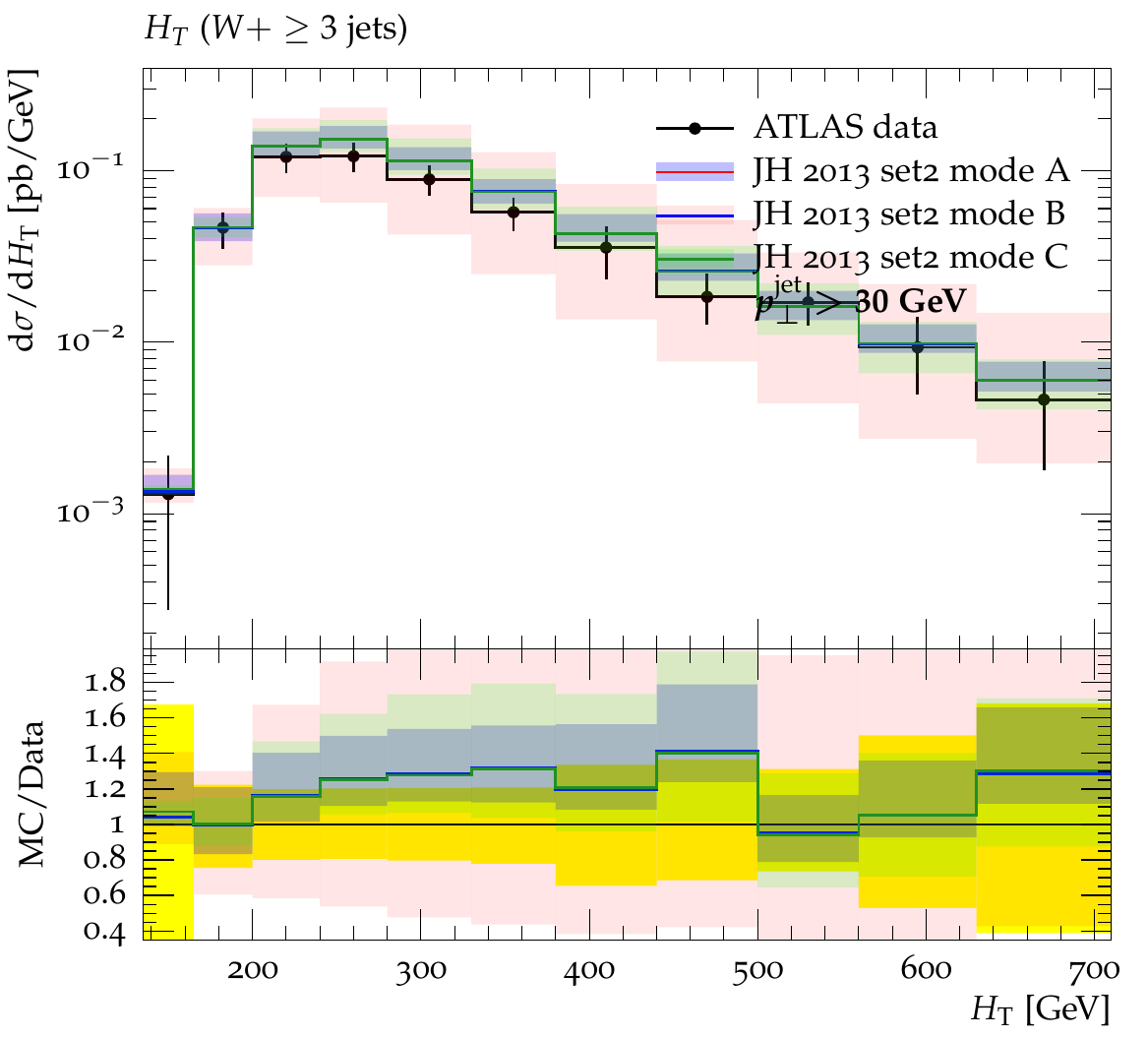}
  \caption{\it Total transverse energy $H_T$ distribution 
  in  final states   
with 
$W $-boson + $n$  jets  at the LHC, for 
 $ n \geq 1 $,  $ n \geq 2 $, $ n \geq 3 $. 
The purple, pink and green bands 
correspond to the mode A, mode B  and mode C,   
 described in~\protect\cite{samantha-w-plb}, to estimate 
 theoretical uncertainties.  The experimental data are 
 from~\protect\cite{atlas-w-jets}, with the  
experimental uncertainty represented by  the yellow band.}
\label{fig:Ht}
\end{figure} 
the  factorization~\cite{hef} 
employed in the previous section is 
 based on the high-energy expansion 
$\sqrt{s} \to \infty$ and it 
   is valid  for    arbitrarily large momentum 
transfers.  
It is designed to take into 
account  TMD physical effects which persist 
at high $p_T$ and can affect 
 final states with high  jet multiplicities~\cite{hj-ang}.  
We  further  note that 
  the implementation~\cite{ref_updfevolv}    
of TMD evolution 
 as  an exclusive  branching  
Monte Carlo  process  
explicitly 
provides, along with the 
parton density,  the detailed 
 structure of the final states.\footnote{This is a 
specific  feature 
of the approach~\cite{ref_updfevolv}   
and 
distinguishes  it both from  ordinary 
parton-shower  Monte Carlo generators and 
from  TMD resummation  programs such as~\cite{landry,guzzi}.}  
Based on these observations, 
Ref.~\cite{samantha-w-plb}  
investigates 
 the applicability of the TMD  theoretical framework 
to  large-$p_T$   processes,  by studying  
 multi-jet final states 
 associated with Drell-Yan (DY) 
production of electroweak gauge bosons at the 
LHC.    

Fig.~\ref{fig:Ht} shows  predictions  
based on the TMD distribution functions discussed 
in the previous section  for   the total 
transverse energy $H_T$ distribution in final 
states with 
$ W $-boson + $n$  jets, with $ n = 1 , 2 , 3 $, at the 
LHC.  For comparison we plot the experimental 
measurements~\cite{atlas-w-jets} 
(jet rapidity $|\eta |  < 4.4$, jet transverse momentum 
$p_T >   30 $ GeV). 
The  main  features of the 
$H_T$ distribution  
 are   described by the 
predictions,   including the case of  higher   jet  multiplicities. The theoretical uncertainties are 
 shown according to three different modes as 
discussed in~\cite{samantha-w-plb}: 
mode A (purple band in the plots) 
includes  uncertainties  due to the  
  renormalization scale, starting evolution scale,  
  and experimental errors, while 
  mode B 
 (pink band in the plots) and mode C (green band in the plots) 
also  include  
 factorization scale 
uncertainties.\footnote{Mode  B is  
to be regarded as the most conservative 
estimate and gives the largest  
uncertainty bands.  
Mode C is  most closely related to 
to the 
estimation of  uncertainties 
 in standard, collinear    calculations, and gives 
bands intermediate between the other two modes. 
See~\cite{samantha-w-plb} for details.}  
The theoretical uncertainties are larger for larger 
$H_T$  (increasing $x$) and,   
at fixed $H_T$,  for higher 
jet multiplicities. 
Ref.~\cite{samantha-w-plb}  also 
 examines  the transverse momentum spectra 
of the individual jets,  and the 
 angular correlations between the jets and of the jets 
with the vector boson. 
Within the uncertainties, a good 
description of the 
measurements~\cite{atlas-w-jets,cms-w-jets} is 
found.   

The main advantage of the exclusive 
TMD approach to $ W $-boson + $n$  jets 
proposed in~\cite{samantha-w-plb}   
is the possibility to construct a formalism 
which  interpolates  from low $p_T$   to 
high $p_T$, and incorporates all-order 
coherence 
effects~\cite{skewang,hj-ang,mw92} 
 associated with multiple 
soft-gluon emission at finite angle, possibly enhanced 
in events with large rapidity 
intervals between final-state particles~\cite{hau-fwd}.   
These effects are  beyond treatments based on 
next-to-leading-order 
perturbation theory matched with collinear parton 
showers~\cite{hoeche13}.  
With automated methods for off-shell multi-leg 
calculations currently being 
developed~\cite{vanham}, the TMD 
 approach could become    a general  tool 
for phenomenological analyses of 
complex  hadroproduction final 
states at the LHC. 

A  general motivation for TMD  approaches  comes 
from the observation of  sizeable   kinematic 
corrections to showering algorithms  due to 
collinearity approximations~\cite{sama1}. 
This, for instance,  underlines the relevance of 
Monte Carlo 
generators~\cite{Jung:2010si,jada09} that  aim to 
go beyond these approximations. 
In the specific case of 
vector  boson +   jets, TMD corrections 
to showering~\cite{eflowproc} may affect 
the interpretation of $W$ + $ 2 j $  
measurements~\cite{ellie} as a signal for double parton 
scattering.  

It should be noted that, unlike the  HERA  
measurements in Sec.~3, 
 the 
 longitudinal momentum fractions  $x$ 
 sampled in the 
 $ W  $-boson + jets cross sections at the LHC are not 
 very   small. Moreover, 
    quark  density   contributions 
  matter at TMD level. 
For these reasons,   $ W  $ + jets   pushes   the limits of  the 
 approach, 
in a manner which can be 
controlled using the estimation of  
theoretical and experimental  uncertainties on 
  TMD  distributions.     
 The results  are  however encouraging, and  
    sufficiently general  to be of interest to  
any  approach  that  employs   TMD formalisms in QCD  
to go beyond fixed-order perturbation theory and 
appropriately    take  account  of   nonperturbative effects. 
This  can  be       relevant  both to  
precision  studies of Standard 
Model physics and to new physics searches 
for which vector 
 boson plus jets production is an important background.
In particular, future applications may  employ  vector boson $ pp $ 
data to advance our knowledge of transverse momentum 
parton distributions. 

% %%%%%%%%%%%%%%%%%%%%%%%%%%%%%%%%%%%%%%%%%%%%%%%%%%%%%%%%%%%%%%%%%%%%%%%%%%%%%%%%

%\newpage 

\begin{footnotesize}

\end{footnotesize}

\end{document}